Article

# Quantum sequel of neural network training

Check for updates

**Hao Zhang** [1] ✉ & **Alex Kamenev** [1,2]

Training of neural networks (NNs) has emerged as a major consumer of both computational and energy resources. Quantum computers were coined as a root to facilitate training, but no experimental evidence has been presented so far. Here we demonstrate that quantum annealing platforms, such as D-Wave, can enable fast and efficient training of classical NNs, which are then deployable on conventional hardware. From a physics perspective, NN training can be viewed as a dynamical phase transition: the system evolves from an initial spin glass state to a highly ordered, trained state. This process involves eliminating numerous undesired minima in its energy landscape. The advantage of annealing devices is their ability to rapidly find multiple deep states. We found that this quantum training achieves superior performance scaling compared to classical backpropagation methods, with a clearly higher scaling exponent (1.01 vs. 0.78). It may be further increased up to a factor of 2 with a fully coherent quantum platform using a variant of the Grover algorithm. Furthermore, we argue that even a modestly sized annealer can be beneficial to train a deep NN by being applied sequentially to a few layers at a time.

Neural networks (NNs)[1–8] have become one of the most transformative technologies, accelerating progress across multiple fields. A key factor in the success of neural networks[7–13] is the development of efficient training methods. Techniques such as stochastic gradient descent[14], backpropagation[15], pre-training[5], ReLU activation[16], residual connections[17] have paved the way for training deep NNs effectively, leading to groundbreaking applications, such as AI. Yet, this progress comes at extraordinary costs, as training state-of-the-art models now routinely requires massive computational and energy resources. Moreover, extrapolating current trends suggests even steeper future costs[18]. This motivates the search for fundamentally different training paradigms.

Meanwhile, quantum technologies have achieved critical milestones in recent years[19–24]. Of particular interest are quantum annealers, analog quantum computing devices engineered to exploit quantum coherent evolution to navigate glassy energy landscapes[25–35]. Modern quantum annealing architectures already approach $10^4$ qubits[33,35], with demonstrated advantage in optimizations[35–38] and quantum dynamics simulations[23,33].

Previous studies have explored the synergy between neural networks and quantum effects, including the use of quantum annealers for training restricted Boltzmann machines and deep belief networks[39–42], the development of variational and circuit-based quantum neural networks[43–47]. Broader perspectives on quantum machine learning can be found in comprehensive reviews[48]. In this paper, we adopt a different perspective, which takes advantage of quantum annealers' capability to rapidly map out low-energy landscapes of (e.g., classical) spin glass models. Here we show how one may take advantage of it to accelerate NNs training. Once trained, such NNs can be deployed on classical hardware, combining the advantages of quantum and classical computing.

The approach is based on the observation that the training process drives NNs through a phase transition from an initial glassy state to a highly ordered trained state[49,50]. The quantum dynamics facilitates such a transition due to its inherent ability to escape local minima[33,38]. It thus allows one to efficiently explore the hierarchical basin structure of the low-energy landscape[38], which appears to be a key for accelerating NNs' training. Being implemented on the D-Wave quantum annealer[32,33], the quantum-assisted training algorithm demonstrates clear practical scaling advantages compared to classical training methods. Moreover, we show that a variant of Grover search algorithm[51], known as amplitude amplification protocol[52], implemented on a fully coherent quantum annealer, can further accelerate NNs' training by potentially doubling the corresponding scaling exponent.

## Methods

### Neural network training

Here we explain the principal idea behind NN training strategy, amenable to be implemented on a quantum annealer. To be specific, consider a NN trained to recognize handwritten images of 0-9 digits. We adopted the architecture and image dataset (a subset of MNIST) as described in ref. 53. The network consists of three layers: one input layer, one hidden layer, and one output layer, see Fig. 1. The input layer contains $28^2 = 784$ units, denoted by $x_a$, where $0 \le x_a \le 1$ and $a = 1, 2, \ldots, 784$, representing grayscale pixel values from handwritten images. The hidden layer consists of 120 units (implemented as qubits), denoted by $s_i^h$, where $i = 1, 2, \ldots, 120$ and $h$ designates the hidden layer, taking values of $\pm 1$ after a measurement of qubits' $z$-component. The output layer contains 40 units (also implemented as qubits), denoted by $s_\alpha^o$, here $\alpha = 1, 2, \ldots 40$ and $o$ stands for the output, also taking values of $\pm 1$ upon $z$-measurement.

[1]School of Physics and Astronomy, University of Minnesota, Minneapolis, MN, USA. [2]William I. Fine Theoretical Physics Institute, University of Minnesota, Minneapolis, MN, USA. ✉e-mail: hao.zhang.quantum@gmail.com

Connections from classical units in the input layer to quantum units in the hidden layer are implemented through local bias fields, $h_i$:

$$h_i[x] = \sum_{a=1}^{784} W_{ia} x_a, \tag{1}$$

where $W_{ia}$ matrix represents coupling strengths between them. Quantum units in the hidden and output layers are connected through couplings $J_{i\alpha}$ in the bare NN Hamiltonian,

$$H_0 = \sum_{i\alpha} J_{i\alpha} Z_i^h Z_\alpha^o + \sum_{i=1}^{120} b_i^h Z_i^h + \sum_{\alpha=1}^{40} b_\alpha^o Z_\alpha^o, \tag{2}$$

where $Z_i^h, Z_\alpha^o$ are Pauli-Z matrices acting on qubits $s_i^h$ and $s_\alpha^o$ respectively, and $b_i^h$ and $b_\alpha^o$ are bias parameters. Note that there are no couplings within the same layer. Hamiltonian of NN with an input image, $x$, is given by:

$$H[x] = H_0 + \sum_{i=1}^{120} h_i[x] Z_i^h, \tag{3}$$

where the argument, $[x]$, emphasizes its dependence on an input training or test image, $x$.

When the neural network has already been trained, it works in the following way. An input image, $x$, to be inferred, is encoded into the bias fields, $h_i[x]$, as in Eq. (1). One looks then for a 160-qubit configuration, $\{s_i^h, s_\alpha^o\}$, providing a ground (or a low-energy) state of the Hamiltonian $H[x]$. Specifically, the output part, $s_\alpha^o$, of the resulting ground state (spin configuration) is used to predict the image class, $\tilde{y} = 0, 1, \ldots, 9$. To this end the 40 output units are grouped into 10 groups by summing every four consecutive output values, forming a 10-component vector. A class predictor, $\tilde{y}$, is given by a position of the maximum component in this vector. This 4-fold redundancy of the output layer serves as a simple majority-rule error-correcting mechanism.

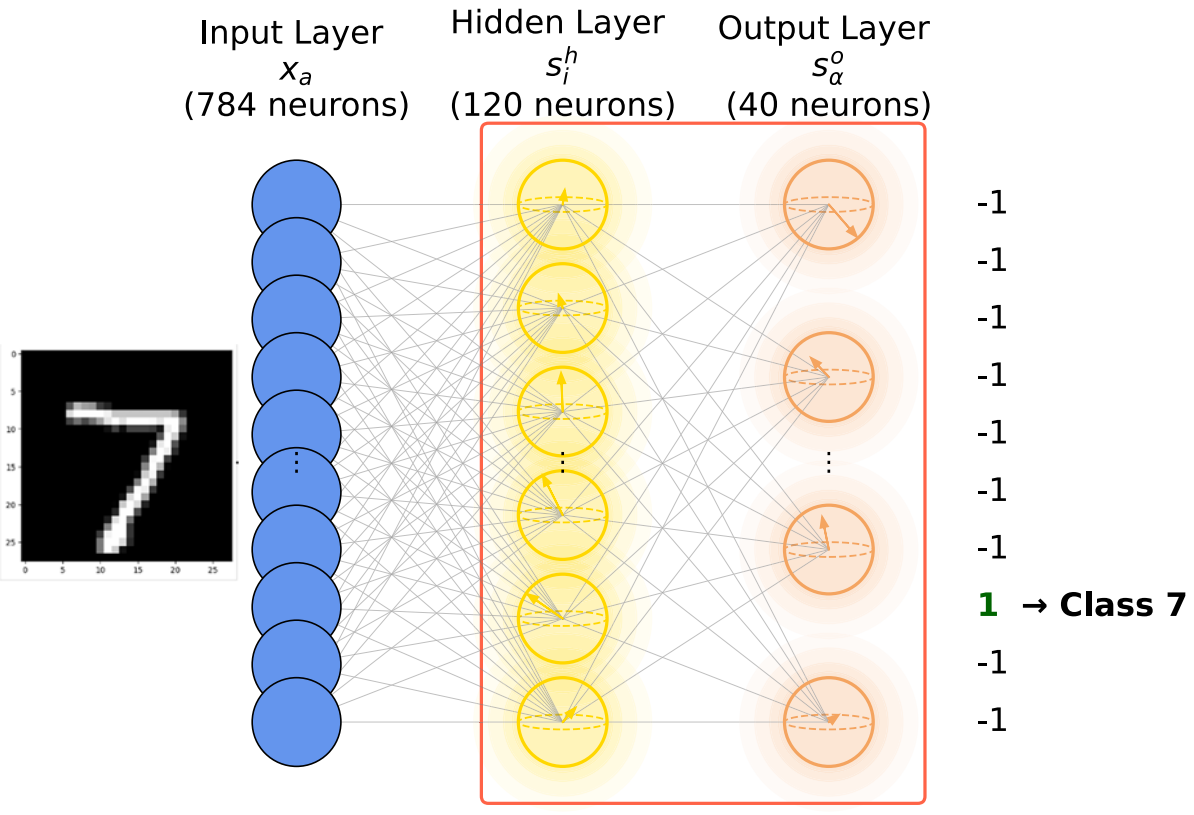

**Fig. 1 | NN architecture implemented on the D-Wave annealer.** The input layer consists of 784 neurons, representing pixels of 28 × 28 MNIST images. The hidden and output layers, comprised of 120 and 40 neurons respectively, are encoded as qubits on the quantum annealer. Each neuron in the output layer represents a digit class (0–9), with the 4-fold error-correcting redundancy. An output layer qubit, measured as $+1$ indicates the corresponding digit class.

As explained below, the fully trained neural network spin-Hamiltonian, $H[x]$, is *not* a glass. Rather, it's energy landscape contains a *single* (in case of the input image resembling one of training images) deep basin of attraction. Therefore, finding its ground state is a computationally easy task, which may be accomplished by, e.g., a simulated classical annealing. On the contrary, the training stage involves dealing with a glassy energy landscape. It can thus significantly benefit from using a quantum annealer, as explained below.

First we describe a training routine, known as *equilibrium propagation*[54], which will be subsequently modified to take advantage of quantum capabilities. To train the NN, it introduces the *nudge* Hamiltonian, $H_N[x, y]$, which depends on a training image, $x$, and its apriori known class $y = 0, 1, \ldots, 9$, as

$$H_N[x, y] = H[x] - \sum_{\alpha=1}^{40} n_\alpha[y] Z_\alpha^o, \tag{4}$$

where $n_\alpha[y]$ is a nudge bias encoding the class, $y$, of each training image, $x$, as $n_\alpha[y] = 1$ for $4y < \alpha \le 4(y + 1)$, and $n_\alpha[y] = -1$ otherwise.

The training process involves multiple updates of the network parameters. Initially, all parameters $W_{ia}$, $J_{i\alpha}$, $b_i^h$, $b_\alpha^o$ are chosen randomly. In our implementation, $W_{ia}$ is drawn from a uniform distribution over $\left[-1/\sqrt{784}, 1/\sqrt{784}\right]$; $J_{i\alpha}$ is drawn from a uniform distribution over $\left[-1/\sqrt{120}, 1/\sqrt{120}\right]$. The biases $b_i^h$, $b_\alpha^o$ are initialized to zero. In each update step, a training image, $x$, belonging to a class $y$, is randomly selected from the dataset. One then looks for two low-energy spin configurations: the first, $\{s_i^h, s_\alpha^o\}$, of the system Hamiltonian $H[x]$, and the second, $\{s_i^{h,N}, s_\alpha^{o,N}\}$, of the nudge Hamiltonian $H_N[x, y]$. The second step may be circumvented, as the strong nudge fixes $s_\alpha^{o,N} = n_\alpha[y]$, while $s_\alpha^{o,h}$ is then simply determined by the sign of the local field as in Eq. (7). The parameters are updated based on the differences between these two spin configurations. Minimization of the loss function leads to the following update rules[54]:

$$\Delta W_{ia} = \delta_W \left( s_i^h x_a - s_i^{h,N} x_a \right), \tag{5a}$$

$$\Delta J_{i\alpha} = \delta_J \left( s_i^h s_\alpha^o - s_i^{h,N} s_\alpha^{o,N} \right), \tag{5b}$$

$$\Delta b_i^h = \delta_h \left( s_i^h - s_i^{h,N} \right), \tag{5c}$$

$$\Delta b_\alpha^o = \delta_o \left( s_\alpha^o - s_\alpha^{o,N} \right), \tag{5d}$$

where $\delta_W$, $\delta_J$, $\delta_h$, and $\delta_o$ are small positive learning rates.

If the low-energy spin configuration sampled from $H[x]$ is identical (or very close) to that from $H_N[x, y]$, it implies that the network already produces the correct output. In this case the parameter do not update. However, if the spin configuration from $H[x]$ differs from the one sampled from $H_N[x, y]$, the update rules penalize this difference by increasing the energy of this incorrect spin configuration. As a result, the likelihood of sampling this spin configuration or its neighbors in the future trials decreases. Through repeated updates, the energy landscapes of $H[x]$ and of $H_N[x, y]$ gradually align. Since the latter is engineered to enforce the output layer units to predict the proper class, $y$, the former inherits the same trait, without having the nudge bias, Eq. (4), (not known apriori for test images). Once this is achieved, the neural network is considered to be trained. Presenting NN with all images from the training set along with corresponding adjustments of the parameters, Eq. (5), is called a training epoch. Number of such epochs needed to fully train NN may range from dozens to thousands.

### Understanding NN training mechanism

To better understand how NN develops its ability to classify images, we utilized D-Wave's quantum annealer to sample low-energy spin configurations of (classical) Hamiltonians (3) and (4) using a quantum annealing protocol (specified below). We used a training set (1000 images, 100 per class) to train the NN for 13 epochs. At various stages of the training process we employed a test set (100 images total, 10 per class) to visualize the energy landscape of NN Hamiltonian, (3). To this end we sampled 100 low-energy spin configurations for each test image, $x$, 10,000 low-energy spin configurations in total. We then applied Multidimensional Scaling (MDS) using the scikit-learn library[55], projecting high-dimensional spin configurations (hypercube with dimensionality $N$ = 160) onto a two-dimensional plane. The MDS algorithm preserves the pairwise Hamming distances as closely as possible in the 2D representation.

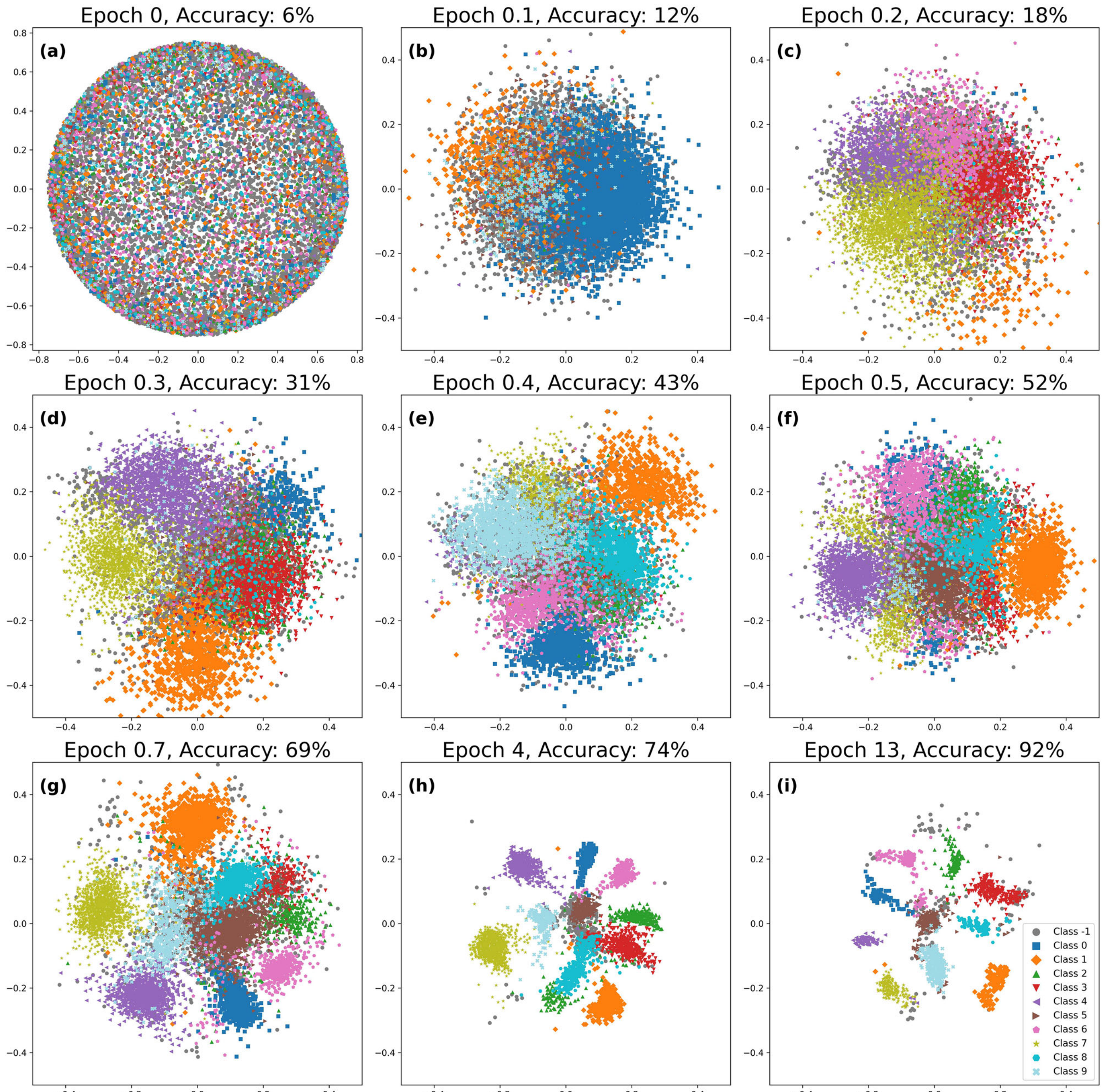

**Fig. 2 | Development of the NN.** The development is visualized in (**a**) to (**i**) through low-energy states of $H[x]$, sampled by the D-Wave annealer. Each subplot corresponds to a different training epoch, with the test accuracy indicated. For each image in the test set (100 images total, 10 per class), 100 low-energy spin configurations were sampled using fixed network weights at a given epoch. These $10^4$ spin configurations are projected from the 160-dimensional hypercube onto 2D using multidimensional scaling (MDS), preserving Hamming distances. Points are color-coded by their output class (0−9); gray points (class−1) indicate unclassified spin configurations. The late stages clearly demonstrate the appearance of energy basins, with 9 of them needed to be cut off for any given training image, $x$.

Figure 2 shows the resulting 2D visualizations for 9 representative instances during the training history. Each point represents one of $10^4$ low-energy spin configurations, color-coded by their output classes (0-9) (with gray points, labeled as class -1, indicating unclassified spin configurations in cases when the output vector does not have a single largest component). The evolution of the low-energy landscape illustrates the development of the NN from a random initialization stage toward a highly structured mature stage. As training progresses, the landscape develops 10 distinct well-separated low-energy basins, representing 10 classes. Correspondingly the NN's recognition accuracy on the test set increases from 6% to 92%. It can be further increased with a few extra training epochs.

Figure 3 gives a more detailed view of this process. The first row displays five examples of training images of the digit 7. Each image is fed into the NN and, as previously explained, low-energy spin configurations of the Hamiltonian $H[x]$ with those images, $x$, belonging to the class $y$ = 7, are sampled and plotted via the 2D MDS visualization. Dots are again color-coded according to their output class, $\tilde{y}$. Notice that the latter may or may not be equal to 7, representing correct or incorrect recognition respectively. The second and third rows display low-energy spin configurations for a poorly- and a well-trained NN. Each column corresponds to the same input image shown in the first row. In the poorly trained network (second row), one observes a mixed distribution of spin configurations associated with multiple output classes. As training progresses, the low-energy spin configurations become more closely spaced and their output class is almost surely $\tilde{y} = 7$.

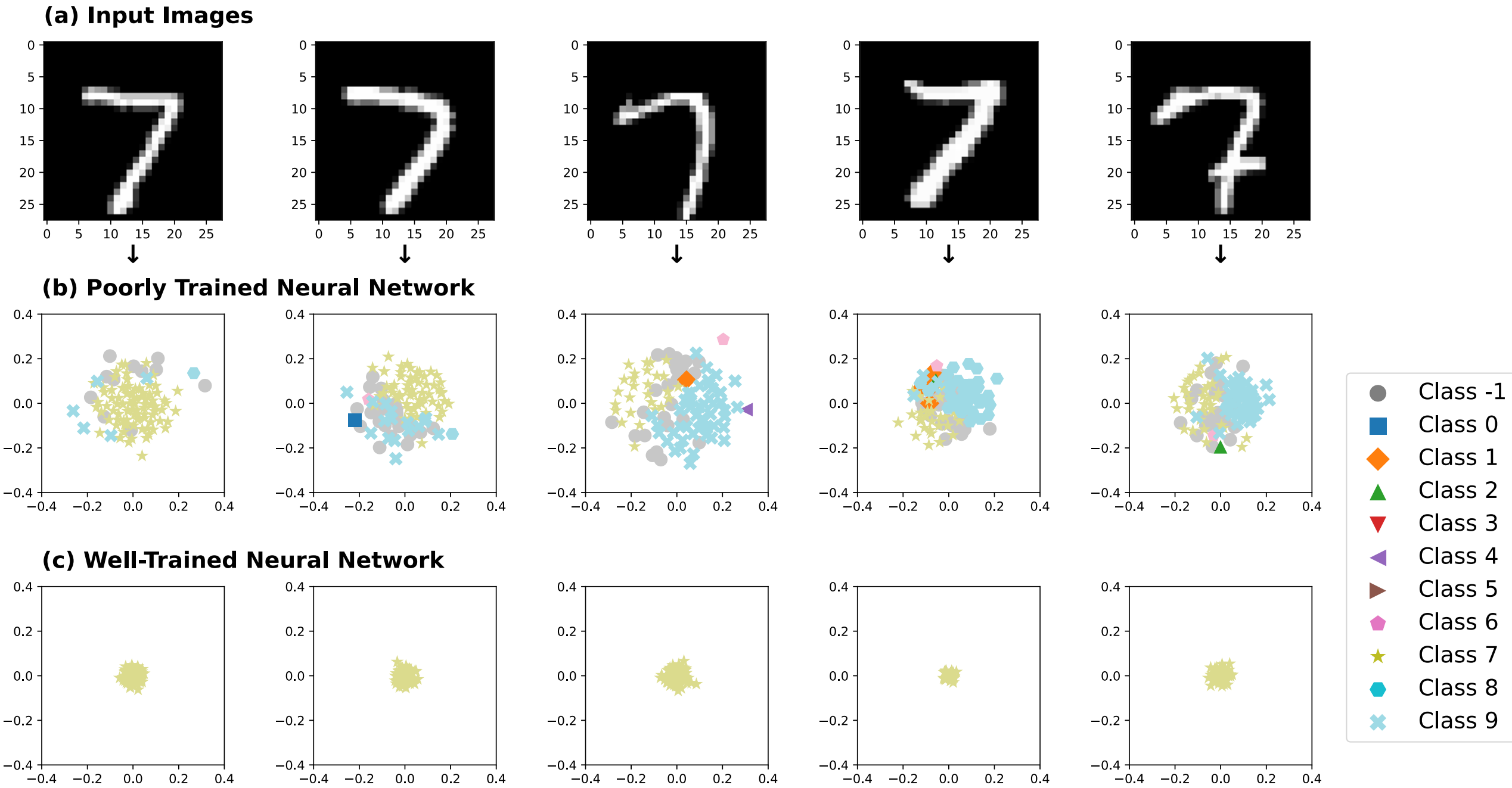

**Fig. 3 | Training NN to recognize digit 7. a** Five representative 28 × 28 test images of digit 7. **b** Corresponding low-energy spin configurations sampled from a poorly trained network's energy landscape. Points are color-coded by their inferred output class ($\tilde{y}$), showing a scattered distribution across several classes. **c** Same test on a well-trained network exhibits tight clustering and uniform output labels ($\tilde{y} = 7$), showing that NN has developed a distinct energy basin for the digit 7. The legend applies to all the panels in (**b**) and (**c**).

The lesson is that the training process can be viewed as a phase transition from a glassy phase (no explicit structure) to an ordered phase (a single deep basin for any image, $x$). (It is extremely important that images with distinct output classes produce such a deep basin in far-apart regions of the hypercube.) Such a transition makes the low energy spin configurations of the NN Hamiltonian loaded with an image, $x$, Eq. (3), to be concentrated within a narrow basin, having the same output class, $\tilde{y} = y$, where $y$ is the class of the image $x$. It achieves this goal by energetically penalizing all states with $\tilde{y} \neq y$. The question is thus if and how a quantum annealer can accelerate such a training transition.

### Quantum training algorithm

The main advantage of quantum hardware is its ability to produce low-energy states (as low as 0.1% above the ground state energy[38]) of Ising-like Hamiltonians, extremely fast (e.g., microseconds in case of D-Wave). Another useful feature is its ability to localize the search to specific parts of the hypercube. The latter capability is provided by the *cyclic annealing* protocol[35,36,38]. Unlike the more traditional forward annealing, it biasses the search to a vicinity of a chosen (so-called reference) state.

The Equilibrium Training algorithm, described above, looks for a low-energy spin configuration, $\{s_i^h, s_\alpha^o\}$, which needs to be sufficiently far from the nudge spin configuration $\{s_i^{h,N}, s_\alpha^{o,N}\}$, to generate evolution of NN parameters according to Eq. (5).

Before presenting the quantum extension of Equilibrium Propagation, it is useful to clarify the role of the so-called wrong spin configurations. For a given training image $x$ with correct label $y$, the nudge Hamiltonian $H_N[x, y]$ enforces an output consistent with $y$. In contrast, the free Hamiltonian $H[x]$ may yield low-energy spin configurations with incorrect outputs $\tilde{y} \neq y$. These are the wrong spin configurations, located in spurious basins in the energy landscape which compete with the correct one. The training updates are precisely driven by the differences between the free spin configurations $\{s_i^h, s_\alpha^o\}$ and the nudge spin configuration $\{s_i^{h,N}, s_\alpha^{o,N}\}$. Whenever a wrong spin configuration is encountered, the update rules penalize it by raising its energy relative to the correct spin configuration, gradually suppressing its future occurrence.

With a quantum annealer, one can efficiently obtain not just a single low-energy spin configuration but a collection of $m \gg 1$ spin configurations, $\{s_i^{h,\gamma}, s_\alpha^{o,\gamma}\}$ with $\gamma = 1, 2, \ldots, m$, for a given training image $x$. By engineering the initial conditions of cyclic annealing, these configurations can be biased to explore wrong basins. To penalize all such configurations simultaneously, one substitutes

$$
\begin{aligned}
s_i^h &\to \frac{1}{m}\sum_{\gamma=1}^{m} s_i^{h,\gamma}, \qquad s_\alpha^o \to \frac{1}{m}\sum_{\gamma=1}^{m} s_\alpha^{o,\gamma}, \\
s_i^h s_\alpha^o &\to \frac{1}{m}\sum_{\gamma=1}^{m} s_i^{h,\gamma} s_\alpha^{o,\gamma}
\end{aligned}
\tag{6}
$$

into Eq. (5). We refer to the resulting training scheme given by Eqs. (5)–(6) as the *quantum propagation* training algorithm.

Figure 2 shows exactly where are those configurations to penalize. They are in the 9 basins with wrong output classes, $\tilde{y} \neq y$. Indeed, Fig. 4 illustrates how performance improves as the number of sampled spin configurations, $m$, increases during a fixed training epoch. All model parameters are initialized identically for each data point. A significant decrease in error rate is observed for $m \lesssim 10$ and plateaus for larger $m$. To further optimize the training one should therefore look for $m = 9$ energy basins (low-energy spin configurations) within wrong basins. To this end one should initialize cyclic annealing from the reference spin configurations which are the nudge spin configurations of the 9 wrong basins. The latter are given by

$$
s_\alpha^{o,\gamma} = n_\alpha[\tilde{y}^\gamma]; \qquad s_i^{h,\gamma} = -\mathrm{sign}\left(\sum_\alpha J_{i\alpha} n_\alpha[\tilde{y}^\gamma] + b_i^h + h_i[x]\right), \tag{7}
$$

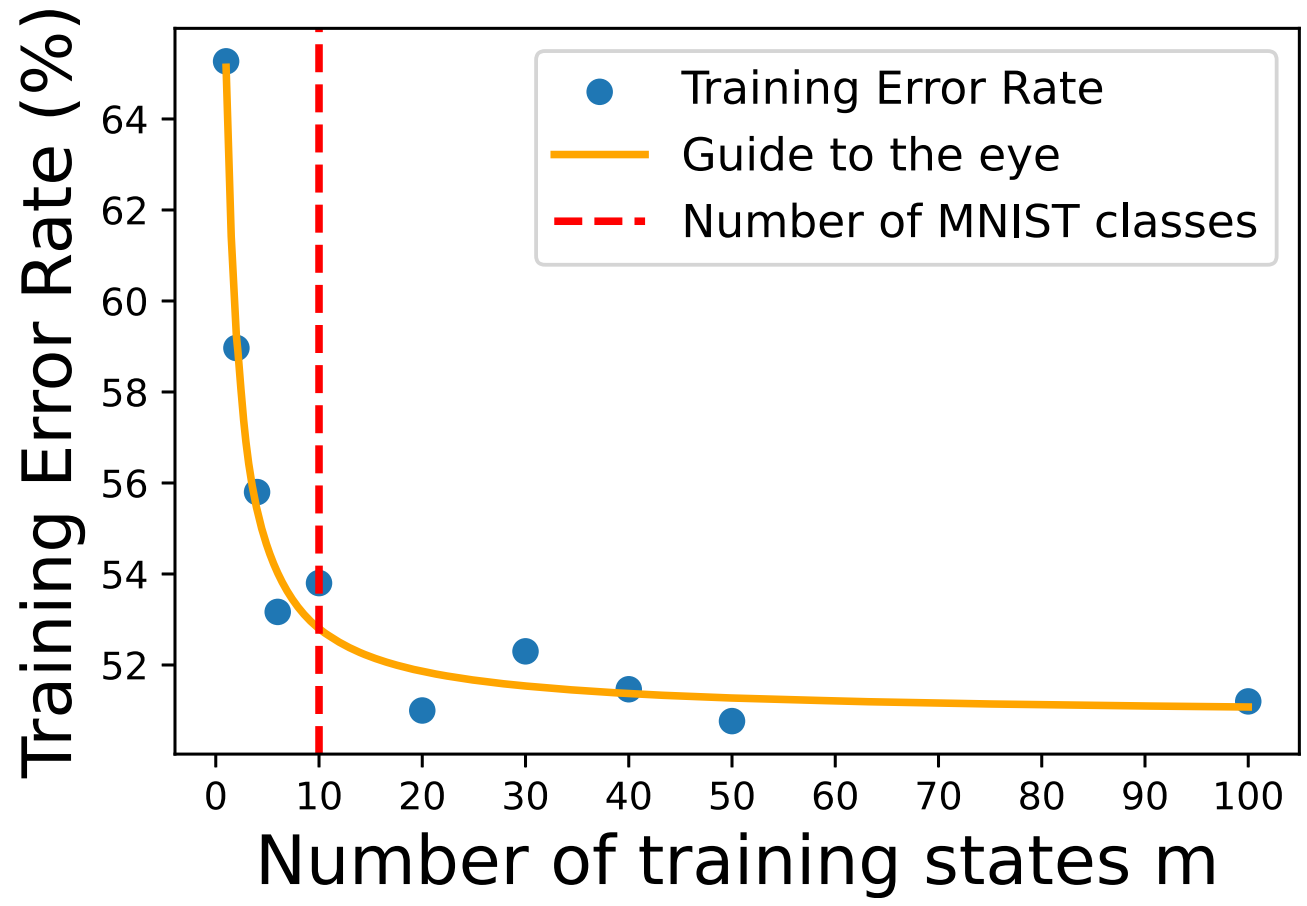

**Fig. 4 | Performance analysis of quantum propagation.** It shows training error rate vs. number of training spin configurations $m$ during the first epoch. The vertical red dashed line marks $m_0 = 10$—the number of MNIST classes. Notable improvement in accuracy is observed when $m \lesssim m_0$. Saturation behavior beyond this point suggests formation of $m_0$ class-labeled basins in the low-energy space.

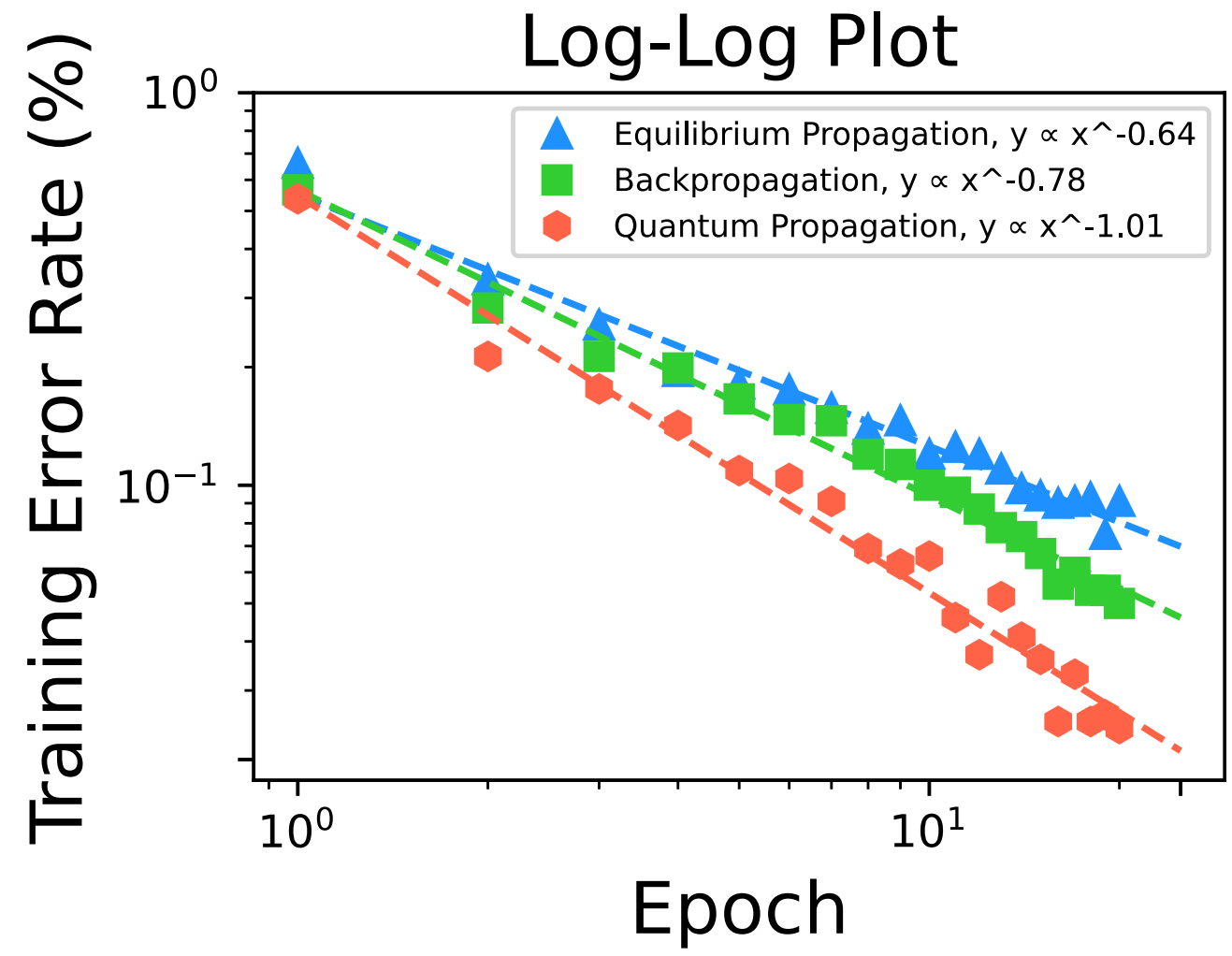

**Fig. 5 | Performance comparison of training methods.** The data is shown on a log-log scale. Quantum Propagation ($m$ = 20, red hexagons, $z$ = 1.01) on D-Wave exhibits advantages scaling performance compared to Backpropagation (green squares, $z$ = 0.78) and Equilibrium Propagation (also on D-Wave) ($m$ = 1, blue triangles, $z$ = 0.64).

where $x$ is the currently presented image with the class $y$ and $\tilde{y}^\gamma \neq y$ are 9 wrong output classes, $\gamma = 1, \ldots 9$.

## Results

### Benchmark

We benchmark the quantum propagation against classical back-propagation. To ensure fairness, both methods employ identical neural network architectures and best-practice settings. Classical training uses real-valued hidden and output units and ReLU activations in the hidden layer. Both methods use simple gradient descent to update parameters. Figure 5 shows the training error rate (1-accuracy) as a function of a number of training epochs in the log-log plot. It exhibits an algebraic relation,

$$\text{error rate} \propto (\#\text{ of epoch})^{-z}, \tag{8}$$

where $z$ is the scaling exponent indicating training efficiency. Blue triangles are data from the equilibrium propagation training, Eq. (5), the corresponding exponent is $z = 0.64$. This is worse than the conventional back-propagation technique (green squares, $z = 0.78$). Red dots represent quantum propagation($m = 20$), exponent $z = 1.01$. The significantly larger exponent $z = 1.01$ for the quantum technique indicates superior performance and efficiency compared to both backpropagation and equilibrium propagation. Importantly, the scaling exponent $z$ serves as a more meaningful performance metric than the raw accuracy alone, as it is unaffected by the computational overhead per parameter update. For a typical number of epochs (100 to 500), a conventional NN training process would require 3 to 4 times the resources to match quantum propagation performance. Additional testing under varied classical conditions did not match Quantum Propagation's efficiency. Specifically, switching to sigmoid activations reduces $z$ to 0.56, and increasing hidden units tenfold (from 120 to 1200 units) only marginally improves $z$ to 0.84.

### Quantum coherent training

Equation (6) reminds an expectation value of corresponding $Z^{h,o}$ operators in a many-body quantum state $|\psi\rangle$, which is a coherent superposition of $m$ bit-string product states. Indeed, if

$$|\psi\rangle = \frac{1}{\sqrt{m}} \sum_{\gamma=1}^{m} |s_i^{h,\gamma}, s_\alpha^{o,\gamma}\rangle, \tag{9}$$

then Eq. (6) takes the form

$$s_i^h \to \langle\psi|Z_i^h|\psi\rangle, \quad s_\alpha^o \to \langle\psi|Z_\alpha^o|\psi\rangle, \quad s_i^h s_\alpha^o \to \langle\psi|Z_i^h Z_\alpha^o|\psi\rangle. \tag{10}$$

At first glance, this does not provide any advantage, since to calculate the expectation values, one should run an annealing protocol multiple times and perform measurements of the corresponding $z$-components. This is exactly what Eq. (6) prescribes to begin with. Yet, there may be a significant benefit hidden here as explained below.

Consider a state $|\psi\rangle$, reached upon a completion of a quantum annealing run, given by $|\psi\rangle = U_{\text{QA}}[x]|\psi(0)\rangle$, where an initial state, $|\psi(0)\rangle$, may be, eg., the $x$-polarized product state, $|\psi(0)\rangle = \prod_{i,\alpha}|+_i, +_\alpha\rangle$. Here $U_{\text{QA}}[x]$ is a unitary quantum evolution operator describing forward annealing with the time-dependent Hamiltonian:

$$H_{\text{QA}}(t) = (1 - s(t)) \sum_{j(\alpha)} X_{j(\alpha)}^{h(o)} + s(t)\, H[x], \qquad 0 \leq t \leq t_f, \tag{11}$$

where $X_{j(\alpha)}^{h(o)}$ are Pauli-$X$'s acting on all qubits, $s(0) = 0$, while $s(t_f) = 1$. The final state, $|\psi\rangle = |\psi(t_f)\rangle$, is a superposition of the two orthogonal components

$$|\psi\rangle = A_y|\psi_y\rangle + A_{\tilde{y}}|\psi_{\tilde{y}}\rangle, \tag{12}$$

where state $|\psi_y\rangle$ has output qubits pointing to the correct class, $y$, of the presented image, $x$, and state $|\psi_{\tilde{y}}\rangle$ has output qubits pointing to all 9 incorrect classes, $\tilde{y} \neq y$. The corresponding amplitudes satisfy $|A_y|^2 + |A_{\tilde{y}}|^2 = 1$.

The goal of the training process is to make $|A_{\tilde{y}}|$ as small as possible. As a result, during advanced training epochs $|A_{\tilde{y}}| \ll 1$. This is a good news for operating NN, but is a very bad news for its further training. Indeed, the "correct" component, $|\psi_y\rangle$, is identical (or very close) to the ground state, $|s_i^{h,N}, s_\alpha^{o,N}\rangle$, of the nudge Hamiltonian. As a result, it does not lead to any improvement of NN parameters, see Eq. (5). It is thus the "wrong" component, $|\psi_{\tilde{y}}\rangle$, measured with the probability $|A_{\tilde{y}}|^2 \ll 1$, which contains all the information about energy basins to be cut off. Therefore the training process may be significantly accelerated if one can use $|\psi_{\tilde{y}}\rangle$ component, instead of $|\psi\rangle$ in Eqs. (10) and (5). The cyclic annealing with reference states, Eq. (7), is not efficient in late epochs, since the basin, defined by the correct class, $y$, is so much attractive that almost every cycle ends up in it, even if the initial reference state is chosen in a basin with $\tilde{y}\neq y$.

Importantly, this can be achieved with the *amplitude amplification* procedure[52,56–58], a generalization of the Grover search algorithm[51], which generates a rotation

$$U_{\mathrm{amp}}|\psi\rangle = |\psi_{\tilde{y}}\rangle. \tag{13}$$

Engineering such a rotation requires order $|A_{\tilde{y}}|^{-1}\gg 1$ applications of $U_{\mathrm{QA}}[x]$ and $U^{\dagger}_{\mathrm{QA}}[x]$ operations, along with a control operation performed on ancilla coupled to the $y$-output qubits ($U^{\dagger}_{\mathrm{QA}}[x]$ operation is achieved by running the annealing in the reversed time direction, $t\to t_f - t$ in Eq. (11)). This should be contrasted with the order $|A_{\tilde{y}}|^{-2}\gg|A_{\tilde{y}}|^{-1}$ annealing runs, required by the straightforward application of Eqs. (10), (12) to observe the $|\psi_{\tilde{y}}\rangle$ component.

As a result, the fully quantum coherent annealing protocol, which incorporates amplitude amplification[52], may significantly further accelerate the training. Namely, it can potentially double the scaling exponent, $z$, in Eq. (8). We have not yet implemented such fully coherent protocol, since the D-Wave device is not expected to maintain coherence over multiple forward and backward runs of the annealing protocol. A smaller version of NN may be tested on a trapped ion platform[57] which exhibits much longer coherence time.

### Training deep neural networks

Finally we discuss a possibility of training deep NNs with $L$ layers with a modest-sized quantum annealer. The neurons in neighboring layers $l$ and $l+1$ are connected through couplings $J_{i_l i_{l+1}}$ in the bare Hamiltonian of a deep NN,

$$H_0 = \sum_{l=1}^{L-1}\sum_{\{i_l,i_{l+1}\}}\left(J_{i_l i_{l+1}}Z^l_{i_l}Z^{l+1}_{i_{l+1}} + b^l_{i_l}Z^l_{i_l}\right) + \sum_{i_L} b^L_{i_L}Z^L_{i_L}, \tag{14}$$

where $Z^l_{i_l}$ are Pauli-$Z$ matrices acting on qubits $s^l_{i_l}$ in layer $l$, and $b^l_{i_l}$ are bias parameters. The last layer, $l=L$, serves as the output layer. To train such deep NN, one looks again for two low-energy spin configurations $\{s^l_{i_l}\}$ of the system Hamiltonian, $H[x]$, and $\{s^{l,N}_{i_l}\}$ of the nudge Hamiltonian, $H_N[x,y]$.

Due to the limited size of quantum annealers, one employs an active-layer sweep procedure. It calls for freezing all qubits along their total $z$-fields, except those in two (or more, depending on the capacity of the annealer) active layers. Starting from randomly initialized parameters, on the forward pass one allows layers $l=1,2$ to be unfrozen and uses the annealer to find their low-energy configuration. Next, with the updated values from layer $l=1$, one freezes all layers except $l=2,3$, and repeat the annealing process for those two layers. Proceeding layer by layer, one sequentially update qubits up to the output layer $L$.

Then, the backward sweep is performed: first, one activates layers $l=L, L-1$ only, and samples their $m$ low-energy spin configurations from $H[x]$ and one from $H_N[x,y]$. These are used to update parameters in layers $l=L, L-1$ via Quantum Propagation rules (5), (6). One then activates layers $l=L-1, L-2$ and repeats updating down to the layer $l=1$. One full forward and backward pass constitutes a single update of the network parameters.

This way one only needs an annealer capable of accommodating two successive layer of a deep NN to take advantage of its ability to rapidly search for multiple low-energy spin configurations. This observation may allow already existing annealers to accelerate training of practical NNs.

## Conclusions

We have shown that even a modest-size quantum annealer can significantly accelerate neural network training. The quantum propagation routine, suggested and implemented here with the D-Wave platform, exhibits a scaling exponent larger than both equilibrium propagation and back-propagation. Moreover, we argued that a fully coherent annealer may further increase the exponent by up to a factor of two. Finally, we discussed a strategy to benefit from quantum annealers with qubit numbers significantly smaller than the total number of neurons.

This work represents an initial step towards implementing quantum devices to accelerate NN training. The favorable scaling observed in our proof-of-principle experiments highlights the potential of quantum devices to surpass conventional training methods. Future directions naturally include larger datasets, deeper architectures, and systematic comparisons with state-of-the-art classical approaches on more challenging tasks. Equally exciting is the prospect of next-generation quantum annealers with longer coherence times, larger qubit counts, and enhanced connectivity, which may enable fully coherent training protocols. Ultimately, one may wonder: if quantum devices prove more effective at navigating complex energy landscapes and training neural networks for increasingly demanding tasks, could they open the door to forms of intelligence beyond what is practically attainable with purely classical methods?

## Data availability

The data that support the findings of this study are provided in Supplementary Data.

## Code availability

The code that supports the findings of this study is available from the corresponding author upon request.

## Acknowledgements

We are grateful to M. Amin, V. Galitski, and A. King for useful discussions. This work was supported by the NSF grant DMR-2338819.

## Author contributions

H.Z. and A.K. conceived the idea. H.Z. performed experiments. H.Z. and A.K. discussed the data and wrote the manuscript.

## Competing interests

The authors declare no competing interests.

## Additional information

**Supplementary information** The online version contains supplementary material available at https://doi.org/10.1038/s42005-025-02384-8.

**Correspondence** and requests for materials should be addressed to Hao Zhang.

**Peer review information** *Communications Physics* thanks the anonymous reviewers for their contribution to the peer review of this work.